\documentclass[12pt]{article}
\usepackage[dvips]{graphicx}
\newcommand{\eq}{\begin{equation}}
\newcommand{\eqn}{\begin{displaymath}}
\newcommand{\en}{\end{equation}}
\newcommand{\enn}{\end{displaymath}}

\def\ds {\displaystyle}

\def\g2{ GeV$^2$}

\def\lsim{\mathrel{\rlap{\lower4pt\hbox{\hskip1pt$\sim$}}
    \raise1pt\hbox{$<$}}}         
\def\gsim{\mathrel{\rlap{\lower4pt\hbox{\hskip1pt$\sim$}}
    \raise1pt\hbox{$>$}}}         
\def\Imag#1{\Im{\rm m}#1}
\def\Real#1{\Re{\rm e}#1}
\def\ie{\hbox{\it i.e. }}

\begin{document}

\rightline{DFTT 30/99}

\rightline{LYCEN 9962}

\bigskip
\begin{center}

{\large {\bf Eikonalization and Unitarity Constraints}}

\bigskip
\bigskip

{\bf P. Desgrolard}({\footnote{E-mail: desgrolard@ipnl.in2p3.fr}}),
{\bf M. Giffon}({\footnote{E-mail: giffon@ipnl.in2p3.fr}}),
{\bf E. Martynov}({\footnote{E-mail: martynov@bitp.kiev.ua}}),
{\bf E. Predazzi}({\footnote{ E-mail: predazzi@to.infn.it}}).

\end{center}

\bigskip
\noindent
($^{1,2}$){\it
Institut de Physique Nucl\'eaire de Lyon, IN2P3-CNRS et Universit\'{e}
Claude Bernard,\\
43 boulevard du 11 novembre 1918, F-69622 Villeurbanne Cedex, France\\}

\noindent
($^3$){\it
N.N. Bogoliubov Institute for Theoretical Physics, National Academy of
Sciences of Ukraine,\\
252143, Kiev-143, Metrologicheskaja 14b, Ukraine\\}

\noindent
($^4$){\it
Dipartimento di Fisica Teorica - Universit\`a di Torino
and Sezione INFN di Torino, Italy}

\bigskip
\bigskip

\noindent {\large{\bf Abstract}}

\noindent An extensive generalization of the ordinary and
quasi-eikonal methods is presented for the $pp$ and $\bar pp$ elastic
scattering amplitudes, which takes into account in a phenomenological
way all intermediate multiparticle states involving the crossing even
and crossing odd combinations of Reggeons. The formalism in this
version involves a maximum of three parameters corresponding to the
intermediate states which are possible in this configuration. The
unitarity restriction is investigated and particular cases are
discussed. An interesting result that emerges concerns the Odderon
trajectory intercept: we find that unitarity dictates that this
quantity {\it must} be below or equal unity unless a very peculiar
equality exists between the coupling of the particles to the Pomeron
and the Odderon.


\section
{\large{\bf Introduction}}

In phenomenological models for elastic scattering of hadrons at
high energy, based on perturbative QCD, the Pomeron is a simple
pole in the complex angular momenta plane, lying at $t=0$ on the
right of $j=1$. This means that the Pomeron trajectory has an
intercept $\alpha_P(0)=1+\delta_P$ with $\delta_P>0$. In this case,
 the Pomeron contribution to the asymptotic
total cross section
$$
\sigma^{tot}_P (s)\ \propto
(s/s_0)^{\delta_P}\ , \quad s_0=1\ {\rm GeV}^2\ ,
$$
results in
a violation of the unitarity, due to the restriction set by the
Froissart-Martin bound~\cite{fm}
$$
\sigma^{tot} (s)\le C\ \ell n^{2}{(s/s_0)}\
, \quad (C \simeq 60\ {\rm mb})\ .
$$
Such a Pomeron - denoted sometimes as {\it supercritical} {\footnote{ The
critical Pomeron with $\delta_P=\delta_{cr}$ is solution of the
Pomeron equation within the Regge Field theory~\cite{rft1,rft2}; in
this theory the supercritical Pomeron has $\delta_P>\delta_{cr}$.}}
- can only be considered as an input or a {\it Born Pomeron} and must
be unitarized. To this aim, the eikonal method ~\cite{col} and its
generalizations are most often used~\cite{ter,um,gmp}.

\smallskip

While the eikonalization procedure is quite standard for one Pomeron
alone (or a group of partners seen as a whole), the situation is more
complicated when, for instance, the Pomeron and others contributors
are considered simultaneously: the problem of the discrimination of
the intermediate states arises. To clarify, let us illustrate the
procedures using the standard or {\it ordinary} eikonal (OE)
model~\cite{col} and the more sophisticated quasi-eikonal (QE)
model~\cite{ter}, before we introduce our generalized eikonal (GE)
model, in the case of the elastic $pp$ and $\bar pp$ elastic
scattering.

Consider the separate form  of the Born amplitude
\begin{equation}
A_{pp, {\rm Born}}^{\bar pp}(s,t)=a_+(s,t)\ \pm a_-(s,t)\ ,
\end{equation}
where the crossing even part takes into account the Pomeron and the
$f-$Reggeon while the crossing odd part takes into account the Odderon and
the $\omega-$Reggeon
\begin{equation}
a_+(s,t)=\ a_P(s,t)+\ a_f(s,t) \ ,\quad
a_-(s,t)=\ a_O(s,t)+\ a_\omega(s,t) \ .
\end{equation}
Let the corresponding crossing even and crossing odd input amplitudes
in the impact parameter $b$-representation be $h_\pm(s,b)$, half the
eikonal function $\chi_{\pm}(s,b) $
\begin{equation}
h_\pm \equiv h_\pm(s,b)\ ={1\over 2} \chi_{\pm}(s,b) \ ={1\over 2s}\
\int_0^\infty \ dq \ q\, J_0(bq)\, a_\pm (s,-q^2)\  .
\end{equation}

\smallskip

The knowledge of the eikonal amplitude in the $b$-representation,
$H^{\bar pp}_{pp}(,s,b)$, in terms of the Born components, $h_\pm
(s,b)$, is at the basis of the eikonalization procedure since in all
eikonal models, once $H^{\bar pp}_{pp}(s,b)$ is known, its inverse
Fourier-Bessel's transform leads finally to the eikonal amplitude in
the $(s,t)$-space, $ A^{\bar pp}_{pp, Eik}(s,t)$, to be used in the
calculation of the observables
\begin{equation}
A^{\bar pp}_{pp, Eik}(s,t)\ =
2s\ \int_0^\infty \ db \ b\, J_0(b\sqrt{-t})\, H^{\bar pp}_{pp}(s,b)\  .
\end{equation}
The standard OE amplitude \cite{col} in the impact parameter
representation, $H_{pp,\!OE}^{\bar pp}(s,\!b)$, is obtained as a sum
of all rescattering diagrams in the approximation for which there are
only two nucleons on the mass shell in any intermediate state
\begin{equation}
H^{\bar pp}_{pp,OE}(s,b)\ =\ {1\over 2i}
\sum\limits_{n=0}^{\infty}\sum\limits_{m=0}^{\infty}
\frac{(2ih_+)^n(\pm 2ih_-)^m}{n!m!}-1=\ {1\over 2i}
\big(       \exp[2i(h_+\pm h_-)]-1  \big) \ ,
\end{equation}
where $-1$ subtracts the term with $n=m=0$. This limitation neglects
the possibility to take into consideration intermediate multiparticle
states. In a QE model \cite{ter} the effect of these multiparticle
states is taken into account generalizing in a phenomenological way
the various exchange diagrams. This is realized introducing an
additional parameter $\lambda$ ($\lambda=1$ corresponds to OE) . The
eikonalized amplitude in the $b$-representation becomes then
\begin{equation}
\begin{array}{rll}
H^{\bar pp}_{pp,QE}(s,b)\ =\ & {1\over 2i\lambda}\left(
\sum\limits_{n=0}^{\infty}\sum\limits_{m=0}^{\infty}
(\lambda)^{n+m}\frac{(2ih_+)^n(\pm 2ih_-)^m}{n!m!}-1\right)\\ =\ &
{1\over 2i \lambda} \left(\exp\big[2i\lambda (h_+\pm
h_-)\big]-1\right)\ .
\end{array}
\end{equation}
However, it is not clear that all intermediate states between two
Pomerons (or one Pomeron and one Odderon {\it etc..}) can be described by
the same parameter $\lambda$. As soon as more Reggeons
come into the game several different parameters could be used.
It follows a much involved formalism. From a
phenomenological point of view, it is necessary to note here, as a
motivation of generalizing actual methods of eikonalization, that the
QE method generally does not lead to a good agreement with $pp$ and
$\bar pp$ data~\cite{dgmp2}.

\smallskip

The present paper is a generalization of the QE method~:
specifically, we go from a one-parameter formalism to a
three-parameters formalism to eikonalize the $pp$ and $\bar pp$
amplitudes. In Sect.~2, as a first step, we deal with the case where
the three parameters entering in the eikonalization procedure obey a
specific relation. In Sect.~3, we discuss the general
three-parameters eikonalized $\bar p p$ and $pp$ amplitudes where the
intermediate states between Pomeron-Pomeron, Pomeron-Odderon and
Odderon-Odderon exchanges are taking into account. General
expressions will be given in simple, closed analytic form. The
constraints arising from unitarity can then be studied in the cases
where the number of parameters is two or three. Especially
interesting in the latter case is that the Odderon intercept must be
below unity.


\section
{\large{\bf Eikonalization of the elastic amplitude with two\\ parameters
describing the intermediate states.}}

\subsection{Amplitudes}

In the following, we simplify the discussion by using 2 {\it
Reggeons} only with a definite $C$-parity, that we call for brevity
the Pomeron ($P$) and the Odderon ($O$). Actually the Pomeron
together with the $f$-Reggeon acts as a first {\it Reggeon} $ \tilde
P=(P+f)$ and the Odderon together with the $\omega$-Reggeon acts as a
second {\it Reggeon} $ \tilde O =(O+ \omega)$. In what follows
(Sects.~2-3) we consider for simplicity the case of $\bar pp$
scattering only (changes for $pp$ are self-evident); later, we return
to the general case by writing the final analytic amplitudes and
discussing the unitarity constraints.

In the QE model [3], we do not discriminate the intermediate states
between $\tilde P-\tilde P, \,\tilde O-\tilde O$ and $\tilde P-\tilde
O$. Releasing this assumption gives rise to a new generalized eikonal
(already denoted as GE) model where the three intermediate states can
be different. We proceed step by step. In the first step, each
intermediate state between the vertices {\it
particle-Reggeon-partic\-le} is correlated with a quantity
 which we denote by
$\root{4}\of {\lambda_i}\, \root{4}\of {\lambda_k} $ ($i,k=+,-$)
where the positive and energy-inde\-pen\-dent parameters $\lambda_i$ and
$\lambda_k$ depend on which { \it Reggeons} are exchanged to
the left and to the right sides of a given diagram. The
(somewhat curious) 1/4 power is chosen so as to
obtain the expressions $\lambda_{\pm}h_{\pm}$ for the final amplitudes

 In terms of these two parameters, we
define the three coefficients
\begin{equation}
\lambda_+= \ C_{PP}, \qquad \lambda_-= \ C_{OO} \qquad \hbox{and} \quad
\lambda_0= \ C_{PO}=C_{OP},
\end{equation}
where the relation
\begin{equation}
 \lambda_0^2=\lambda_+\lambda_- \
\end{equation}
has been assumed{\footnote {Strictly speaking, we should have
used $C_{\tilde P \tilde P}$ etc. instead of $C_{PP}$ etc. in
(7) to recall that $\tilde P= P+f$ and $\tilde O =P+\omega$
but we decided not to do it; we will ignore this formal
complication throughout the paper.}}.

Such a procedure, roughly speaking, mimics a situation where the {
\it particle\--Pome\-ron\--particle} and the { \it
particle-Odderon-particle} amplitude vertices ($g_+$ and $g_-$) are
re\-sca\-led by {\it a priori} different positive constants ($\root
\of {\lambda_+}\ {\rm and} \ \root \of {\lambda_-}$). This means
that, formally, we change the {\it coupling constants} of the
{\it Reggeons} as
$$
g_+\rightarrow \sqrt{\lambda_+}g_+\qquad \mbox {for $ pPp$-vertex},
 $$
 $$
g_-\rightarrow \sqrt{\lambda_-}g_-\qquad \mbox {for $ p\,Op$-vertex}.
$$
However, this implies that for each diagram, we have now extra
multipliers originated from the vertices of the extreme left and
right {\it Reggeons}. It is necessary to divide the contributions of these
diagrams by well defined factors. There are, in fact, three kinds of
rescattering diagrams with $n$ Pomerons and $m$ Odderons
that one should consider

- first kind: both the extreme left and the extreme right
{\it Reggeons} are $P$

- second one: both the extreme left and the extreme right
{\it Reggeons} are $O$

- third one: the extreme left (right) {\it Reggeon} is $P$ while the
extreme right (left) {\it Reggeon} is $O$ (or {\it vice versa)}.
Consequently, the diagrams of the first type must be divided by
$\lambda_+$, the second ones by $\lambda_-$ and the third ones by
$\sqrt{\lambda_-\lambda_+}$.

Consider now each kind of diagrams separately, beginning with the
first kind (illustrated in the Fig.~1). We work out the case for
$\bar p p$ but the same procedure applies to the $pp$ case.

                     \vfill\eject

\begin{center}
\includegraphics*[scale=0.6]{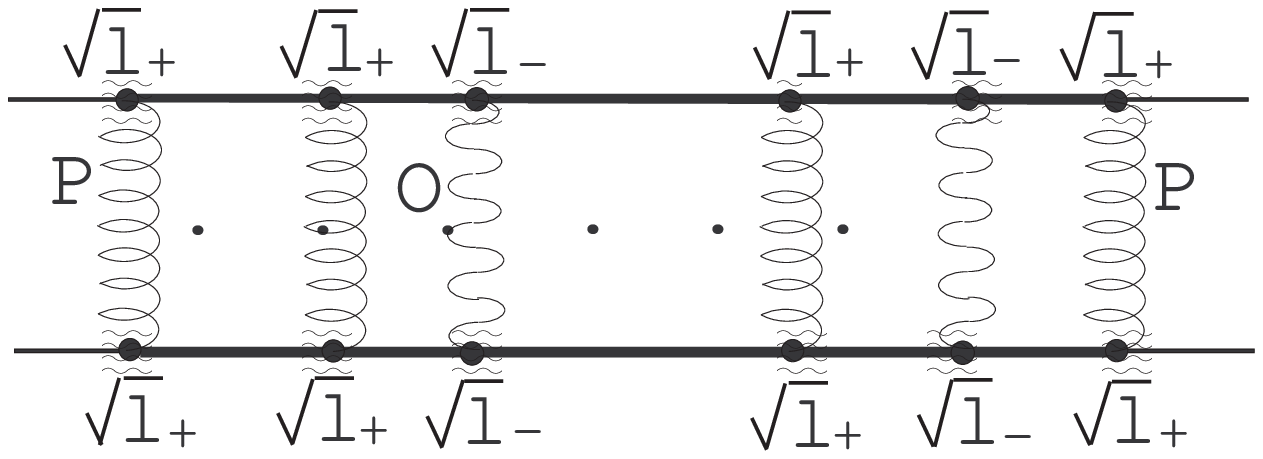}
\vskip 0.3cm
{\bf Fig.~1.} A typical rescattering diagram with a Pomeron at both ends.
\end{center}

\noindent
 the contribution of all the
diagrams (consisting in $n$
Pomerons and $m$ Odderons) is the following
\begin{equation}
H[PP]=\frac{1}{2i\lambda_+}\sum\limits_{n=2}^{\infty}
\sum\limits_{m=0}^{\infty}
(2i\lambda_+h_+)^n(2i\lambda_-h_-)^m\frac{(n+m-2)!}{(n-2)!m!(n+m)!}\ .
\end{equation}
The factor $\frac{1}{(n+m)!}$ takes into account the total number of
$n+m$ { \it Reggeons}; the other factor $\frac{(n+m-2)!}{(n-2)!m!}$
(number of ways to choose $m$ Odderons and $n-2$ Pomerons
from the $n+m-2$ {\it Reggeons}) accounts for all permutations of non
identical Pomerons and Odderons keeping two
Pomerons at the left and right ends of diagrams.

\noindent
Similarly for the
other contributions, one obtains
\begin{equation}
H[OO]\ =\ \frac{1}{2i\lambda_-}\sum\limits_{n=0}^{\infty}
\sum\limits_{m=2}^{\infty}
(2i\lambda_+h_+)^n(2i\lambda_-h_-)^m\frac{(n+m-2)!}{n!(m-2)!(n+m)!}\ ,
\end{equation}
\begin{equation}
\begin{array}{rcl}
& H[PO]=H[OP]\ =\hfill \\
&\ds\frac{1}{2i\sqrt{\lambda_+\lambda_-}}\sum\limits_{n=1}^{\infty}
\sum\limits_{m=1}^{\infty}
(2i\lambda_+h_+)^n(2i\lambda_-h_-)^m\frac{(n+m-2)!}{(n-1)!(m-1)!(n+m)!}\
.
\end{array}
\end{equation}

Defining
\begin{equation}
 H^{\bar pp}(s,b)\ =\ h_+\ +\ h_- \ +\ H[PP]+H[OO]+2 H[PO],
\end{equation}
the $\bar p p$ and the $ pp$ two-parameters eikonalized amplitudes in the
GE case in the impact parameter representation take the final form
(see Appendix A)

$$ H^{\bar pp}_{pp, GE}(s,b)\ =\ h_+\pm h_- $$
\begin{equation}
+\ \left(\frac{h_+\sqrt{\lambda_+}\pm h_-\sqrt{\lambda_-}}
{h_+\lambda_+\pm h_-\lambda_-}\right)^2
\left(\frac{e^{2i(h_+\lambda_+\pm h_-\lambda_-)}-1}{2i}
- (h_+{\lambda_+}\pm h_-{\lambda_-})               \right)\       .
\end{equation}
Recall that, as stressed previously, this result is obtained in the
case of 2 { \it Reggeons} irrespective of whether or not we consider
them as being $P$ and $O$ only or whether we have grouped them
together into the crossing even $ \tilde P= P+f$ and the crossing odd
$ \tilde O=O+\omega$ combinations. Accordingly, we have the
definitions (1-3), if the intermediate states "depend" only on the
parity but not on the specific { \it Reggeon} ($P,f$ or $O,\omega $).

A similar compact formula can be obtained in the case of the so-called
Generalized U-Matrix model [5]. With the same notations one obtains
\begin{equation}
H^{\bar pp}_{pp, GUM}(s,b)=\frac{(h_+\pm h_-)\mp \ds{1\over
2i}h_+h_-\Bigl (\sqrt{\lambda_+}-\sqrt{\lambda_-}\,\Bigr
)^2}{1-2i(h_+\lambda_+ \pm h_-\lambda_-)}\ .
\end{equation}

Actually, here we confine ourselves to the simplest case when the
input amplitudes are purely elastic. In the most general case, we
should also introduce other amplitudes corresponding to different
effective couplings at the extreme left and right ends of diagrams
when the initial (or final) state in the corresponding vertex is not
a single proton (similarly for the "internal amplitudes" inside the
$n$-Reggeons diagrams). These new types of amplitude are the analog
of those considered in diffraction dissociation (with not too high
effective masses). By integration and summation over many
intermediate states, new amplitudes would be derived by modifying
appropriately each $h_i$ and $\lambda_i$. The important difference
would be that these new amplitudes would have different energy
independent parts in their slopes (in agreement with the data) but,
for large $s$, would reduce to the present amplitudes. We will not
consider this additional complication here.

\subsection{Unitarity constraints}

The unitarity inequality
\begin{equation}
|H^{\bar pp}_{pp}(s,b)|\leq 1\
\end{equation}
restricts the admitted values for the parameters $\lambda_+$ and
$\lambda_-$.

In Appendix B we briefly discuss the framework of the input
(or {\it Born}) amplitudes to be used in the general scheme
in both $(s,t)$ and $(s,b)$-representations obtained the one from the
other via a Fourier-Bessel transform. From the formulae of
Appendix B, valid at high energy if the secondary Reggeons are neglected,
one sees that the exponential term in (13) can
be neglected because $h_+$ becomes mainly imaginary and its
modulus increases with the energy. Thus, keeping
the main and the next orders in $h_-/h_+$, we obtain
 $$
 \bigg |h_+\pm h_--h_+\bigg
(1\pm 2\sqrt{\frac{\lambda_-}{\lambda_+}}\frac{h_-}{h_+}
\mp \frac{\lambda_-h_-}{\lambda_+h_+}+\frac{1}{2i}
\frac{1}{h_+\lambda_+}\bigg )\bigg |\le 1\\ ,
$$
which is the same as
\begin{equation}
\bigg
|\frac{1}{2i\lambda_+}\mp h_-\bigg(1-\frac{\sqrt{\lambda_-}}
{\sqrt{\lambda_+}}\bigg )^2\bigg |\le1\ .
\end{equation}
If the crossing even (or Pomeron) and the crossing odd (or Odderon)
trajectories are written as
\begin{equation}
\alpha_\pm (t)=1+\delta_\pm +\alpha'_\pm t\ ,
\end{equation}
the inequality (16) can be satisfied either if{\footnote {When
$s\rightarrow\infty$ and $b$ is inside the interaction radius
($0\leq b^2 < 4\alpha'_+\delta_+\ln^2 s=R^2$ [10,11],
 $|h_-|\rightarrow \infty$ (see Appendix B).}}
\begin{equation}
\lambda_+ = \lambda_-\ ,\quad \mbox{and}\quad \lambda_+ \geq 1/2\
,\quad \delta_-\geq 0\
\end{equation}
or, when $\lambda_+ \not=\lambda_-$, if{\footnote {No restriction is
found on $\lambda_-$, because, in this case, $|h_-| \rightarrow 0$
when $s\rightarrow \infty$ and $0 \leq b^2 < R^2$ ( see Appendix
B).}}
\begin{equation}
\lambda_-\quad \mbox{is arbitrary},\quad\lambda_+\geq 1/2
\ ,\quad\delta_-\leq
0\ ,
\end{equation}

Similar results can be obtained for the two-parameters Generalized
$U-$Matrix model (GUM) [5] in which case from (15) unitarity requires

\begin{equation}
\bigg |\frac{(h_+\pm h_-)\mp \ds{1\over
2i}h_+h_-\Bigl (\sqrt{\lambda_+}-\sqrt{\lambda_-}\,\Bigr
)^2}{1-2i(h_+\lambda_+ \pm h_-\lambda_-)}\bigg |\le1\ .
\end{equation}
One sees at once that the second term of the numerator is dangerous
for the unitarity restriction (15) because it dominates at $s\to
\infty$ if $\delta_+>0$ and $\delta_->0$ and grows faster than the
denominator. In this case, there are three solutions to prevent violation of
unitarity
\eqn
\begin{array}{cl}
  (i) \qquad\qquad\qquad & \lambda_+=\lambda_-\ , \\
  (ii) \qquad\qquad\qquad & \delta_-\leq 0\ , \\
  (iii) \qquad\qquad\qquad & \delta_+\leq 0\ .
\end{array}
\enn
Notice that, compared to the GE, in the GUM we have a third
solution (iii) which is technically possible when $\delta_- >
\delta_+$ in spite of its apparent non realistic aspect [5]. To see
how the above solutions arise we note that when $\delta_+ > \delta_-$
and $|h_+|\rightarrow \infty$, the only way to satisfy (20) is when
$\lambda_+ = \lambda_-$ and $\lambda_+\geq 1/2$ ( the other two
solutions arise similarly, see [5]).

Summarizing, we find that
the two-parameters GE model fulfills
the unitarity requirement (15) when

\noindent
\hskip 1cm
({\it i}) either it reduces to the QE ($U-$matrix) model with one
parameter ($\lambda=\lambda_+=\lambda_-$), with
\begin{equation}
\lambda\geq 1/2\ ,\quad \delta_{\pm}> 0\ ,
\end{equation}
\hskip 1cm
({\it ii}) or it satisfies the limitations
\begin{equation}
\lambda_+\geq 1/2\ ,\quad \mbox{if}\quad \delta_-\leq 0,\quad
\delta_+ >0\ ,
\end{equation}
\hskip 1cm
({\it iii}) or, finally, it obeys
\begin{equation}
\lambda_-\geq 1/2\ ,\quad \mbox{if}\quad \delta_+\leq 0, \quad  \delta_- >0
\quad (\mbox{only for the $U-$matrix})\ .
\end{equation}
These conditions have to be combined with the well-known
Pomeron/Odderon hierarchy \cite{FFKT,M}
$$
\delta_+\ge\delta_-\ ,
\quad \alpha'_+\ge\alpha'_-\ ,
$$
for the case (21) and
$$
\delta_-\ge\delta_+\ , \quad \alpha'_-\ge\alpha'_+\ ,
$$
for the $U-$matrix case (23) where $\alpha'_\pm$ are the slopes of
the linear trajectories for the Pomeron (Odderon). These conclusions
are shown in Table 1, where in order to emphasize the differences
between the various classes of models, we give also the asymptotic
behaviour of $\sigma_{tot}^{pp,\bar pp}$ (see~\cite{gmp}).

\def\init{\tabskip 0pt\offinterlineskip}
\def\crr{\cr\noalign{\hrule}}

$$\vbox{\init\halign to 14.cm{
\strut#&\vrule#\tabskip=1em plus 2em&

\hfil$#$\hfil&
\vrule$\,$\vrule#&
\hfil$#$\hfil&
\vrule#&
\hfil$#$\hfil&
\vrule#&
\hfil$#$\hfil&
\vrule#\tabskip 0pt\crr
& &            &&         &&      && U-{\rm matrix\ only} &\crr
& &\lambda_+    &&=\lambda \geq 1/2  && \geq 1/2 && any &\cr
      & &\lambda_-    &&=\lambda \geq 1/2  &&  any    && \geq 1/2&\crr
& &\delta_+ && \geq \delta_-&& >0 &&\leq 0 &\cr & &\delta_- && >0
&&\leq 0 && > 0 &\crr & &\alpha'_+ && \geq\alpha'_- &&\geq\alpha'_-
&&\leq\alpha'_- &\crr
& & && && && &\cr
 & & \sigma^{pp,p\bar p}_{tot}\propto
 && \alpha'_+\delta_+\ell n^2 s
&&\alpha'_+\delta_+\ell n^2 s&&\alpha'_-\delta_-\ell n^2 s&\crr }}$$

\medskip

\centerline {\bf Table 1 } Summary of unitarity constraints of the GE
and GUM procedures with 2 $\lambda$ parameters. Besides the
limitations on the $\lambda_\pm$ parameters, we show those on the
trajectories in the different cases together with the asymptotic
cross-section
 when the Pomeron (Odderon) dominates.


\section
{\large{\bf Eikonalization of the elastic amplitude with three parameters}}

\subsection{ Contributions to the amplitudes }

We consider now the more general case where (8) is not valid
$$
\lambda_0^2\neq \lambda_+\lambda_-\ .
$$
In this case, three different coefficients
$\lambda_0,\lambda_+,\lambda_-$ should be considered. \noindent
Again, as in Section 2, (let us remind P,(O) means here in fact
$\tilde P, (\tilde O)$) we deal with three types of diagrams and
corresponding terms of the amplitudes $H[PP], H[OO]$ and $H[PO]$. We
start with the first term $H[PP]$ (see Fig.~2). There are $n$
Pomerons and $m$ Odderons, distributed in $i$ cells. The maximal
value of $i$ is \eqn
\begin{array}{ccc}
  n-1 & \qquad \mbox{if}\qquad & m>n-1 \\
  m & \qquad \mbox{if}\qquad & m\leq n-1
\end{array}
\enn
Suppose that one Odderon is in each of the $i$ cells. Then, the
number of ways to choose $i$ cells from the $n-1$ cells is
$$
{n-1\choose i} =\frac{(n-1)!}{i!(n-1-i)!}\ .
$$
The number of ways to distribute the remaining $m-i$ Odderons into
the $i$ cells is
$$
{(m-i)+i-1\choose m-i }=
{m-1\choose i-1}=\frac{(m-1)!}{(i-1)!(m-i)!}\ .
$$
Therefore the total number of diagrams with $n$ Pomerons and $m$
Odderons (distributed in $i$ cells among the Pomerons) is
$$
{n-1\choose i}{m-1\choose i-1}\ .
$$
\begin{center}
\includegraphics*[scale=0.8]{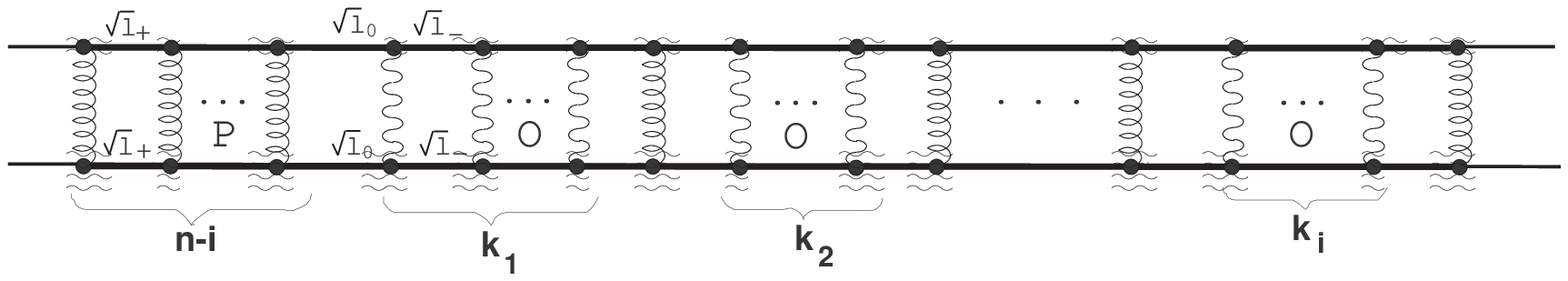}
\end{center}
{\bf Fig.~2.} A typical diagram with a Pomeron at both
ends in the case when the Odderons are grouped in $i$ cells .

\smallskip

For each diagram, as exemplified in Fig.2, we have a certain overall
factor made of powers of $ \lambda_+$, $\lambda_-$ and $\lambda_0$
which we now proceed to calculate. If there are $k_1,k_2,...,k_i$
Odderons in the $1^{\rm st},2^{\rm nd},...,i^{\,\rm th}$ cell, the
$\ell^{\rm th}$ cell contributes a the factor $\lambda_-^{k_\ell-1}$.
Consequently all the cells give the factor
$$
\lambda_-^{k_1-1}\lambda_-^{k_2-1}\cdots \lambda_-^{k_i-1}=
\lambda_-^{k_1+k_2+\cdots +k_i-i}=\lambda_-^{m-i}\ ,
$$
because $k_1+k_2+\cdots +k_i=m.$ In addition, each cell gives
$\lambda_0^2$, therefore for all the cells we have $(\lambda_0^2)^i$.
Furthermore, the number of cells without Odderons inserted is $n-1-i$
and this gives a factor $\lambda_+^{n-1-i}.$ Thus, the total factor
for this diagram is
$$
(\lambda_0^2)^i\lambda_-^{m-i}\lambda_+^{n-1-i}=
\frac{1}{\lambda_+}\bigg (\frac{\lambda_0^2}{\lambda_+\lambda_-}\bigg )^i
\lambda_+^n\lambda_-^m\ .
$$
Besides this, the factor $2ih_+$ corresponds to each Pomeron
amplitude and the factor $2ih_-$ corresponds to each Odderon
amplitude and, remember, there are, altogether n Pomerons and m
Odderons amplitudes.

Summing everything up, the contribution to the rescattering amplitude
of all diagrams with Pomerons at the left and
right ends of each diagram has the following form
\begin{equation}\label{24}
\begin{array}{rcl}
2i\lambda_+H[PP]&=&
\sum\limits_{n=2}^\infty\sum\limits_{m=1}^{n-1}\sum\limits_{i=1}^{m}
\frac{1}{(m+n)!}{n-1\choose i}{m-1\choose i-1}\bigg (\frac{\lambda_0^2}
{\lambda_+\lambda_-}\bigg )^i(2ih_+\lambda_+)^n(2ih_-\lambda_-)^m \\
\displaystyle &+&
\sum\limits_{n=2}^{\infty}\sum\limits_{m=n}^{\infty}\sum\limits_{i=1}^{n-1}
\frac{1}{(m+n)!}{n-1\choose i}{m-1\choose i-1}\bigg (\frac{\lambda_0^2}
{\lambda_+\lambda_-}\bigg )^i(2ih_+\lambda_+)^n(2ih_-\lambda_-)^m\\
\displaystyle &+&
\sum\limits_{n=2}^{\infty}\frac{1}{n!}(2ih_+\lambda_+)^n\ ,
\end{array}
\end{equation}
the last term taking into account the diagrams without Odderons.

\noindent
The same form but with the replacements $h_+\longleftrightarrow h_-$ and
$\lambda_+\longleftrightarrow \lambda_-$ holds for $H[OO]$.
Similarly, one obtains for the contribution of diagrams with a
Pomeron at one end and an Odderon at the other one
\begin{equation}\label{25}
\begin{array}{rcl}
 2i\lambda_0H[PO]&=&\sum\limits_{n=1}^{\infty}\sum\limits_{m=1}^{n}
\sum\limits_{i=1}^{m}
\frac{1}{(m+n)!}{n-1\choose i-1}{m-1\choose i-1}\bigg (\frac{\lambda_0^2}
{\lambda_+\lambda_-}\bigg )^i(2ih_+\lambda_+)^n(2ih_-\lambda_-)^m \\ &+&
\sum\limits_{n=1}^{\infty}\sum\limits_{m=n+1}^{\infty}\sum\limits_{i=1}^{n}
\frac{1}{(m+n)!}{n-1\choose i-1}{m-1\choose i-1}\bigg (\frac{\lambda_0^2}
{\lambda_+\lambda_-}\bigg )^i(2ih_+\lambda_+)^n(2ih_-\lambda_-)^m.
\end{array}
\end{equation}

\subsection{Analytical form of the amplitude}

It is somewhat surprising that one can obtain a compact analytical
form of the total amplitude. To do this, we begin by summing over $i$
in the previous expressions. Introducing
\begin{equation}
z={\frac{\lambda_0^2}{\lambda_+\lambda_-}}\ ,
\end{equation}
and setting $N=m$ (or $N=n-1$) in (24), we obtain
$$
\begin{array}{rcl}
&\ds\sum\limits_{i=1}^{N}{n-1\choose i}{m-1\choose i-1}z^i \
=\displaystyle{{z\over m}}\ {d\over dz}\sum\limits_{p=1}^m\
\displaystyle{{z^p\over(p\,!)^2}}\ (1-n)_p (-m)_p\\ &=\
\displaystyle{z\over m} {d\over dz}\ _2F_1(-m, 1-n;1;z)=
\displaystyle{z(n-1)} _2 F_1(1-m,2-n;2;z) \ ,
\end{array}
$$
where $(a)_b=\Gamma(a+b)/\Gamma(a)$ is the Pochhammer symbol and
the definition of the hypergeometric function $_2F_1$ has been used
in~\cite{pru}.
Similarly, setting $N=m$ (or $N=n$) in (25), we get
$$
\sum\limits_{i=1}^{N}{n-1\choose i-1}{m-1\choose i-1}z^i \
= z \ _2F_1 (1-m,1-n;1;z)\ .
$$
Substituting these results we get
\begin{equation}
2i \lambda_+H[PP]=z\ \sum\limits_{n=2}^{\infty}\sum\limits_{m=1}^{\infty}
{x^n y^m\over(n+m)!}\ (n-1) \ _2F_1(1-m,2-n;2;z) \ +\ e^x-x-1\ ,
\end{equation}
and
\begin{equation}
2i \lambda_0H[PO]=z\ \sum\limits_{n=1}^{\infty}\sum\limits_{m=1}^{\infty}
{x^n y^m\over(n+m)!}\ _2F_1(1-m,1-n;1;z) \ ,
\end{equation}
where
\begin{equation}
x=2i \lambda_+ h_+\ ,\quad y=2i \lambda_- h_-\ .
\end{equation}
We obtain
$\lambda_-H[OO]$ from $\lambda_+H[PP]$ with the replacement
$x \longleftrightarrow y$.

\medskip

To perform the remaining summations, we use the following integral
representation which defines $_2F_1(a,b;c;z)$ as an analytic function
in the $z$-plane, by means of the contour integral~\cite{eder}
\begin{equation}\label{30}
_2F_1(a,b,c ;z)=\frac{-i\Gamma(c)\exp{(-i\pi b)}}
  {2\Gamma (b)\Gamma  (c-b)\sin{\pi b}} \oint\limits_{C}\,dt\, t^{b -1}
  (1-t)^{c-b-1}(1-zt)^{-a}\, ,
\end{equation}
where
$$
\Real{c}\ >\ \Real{b}\ , \quad |arg(-z)|<\pi\ , \quad b\ne 1,2,3,...\ ,
$$
and where the integration contour is defined in Fig.~3.

\begin{center}
\includegraphics*[scale=0.5]{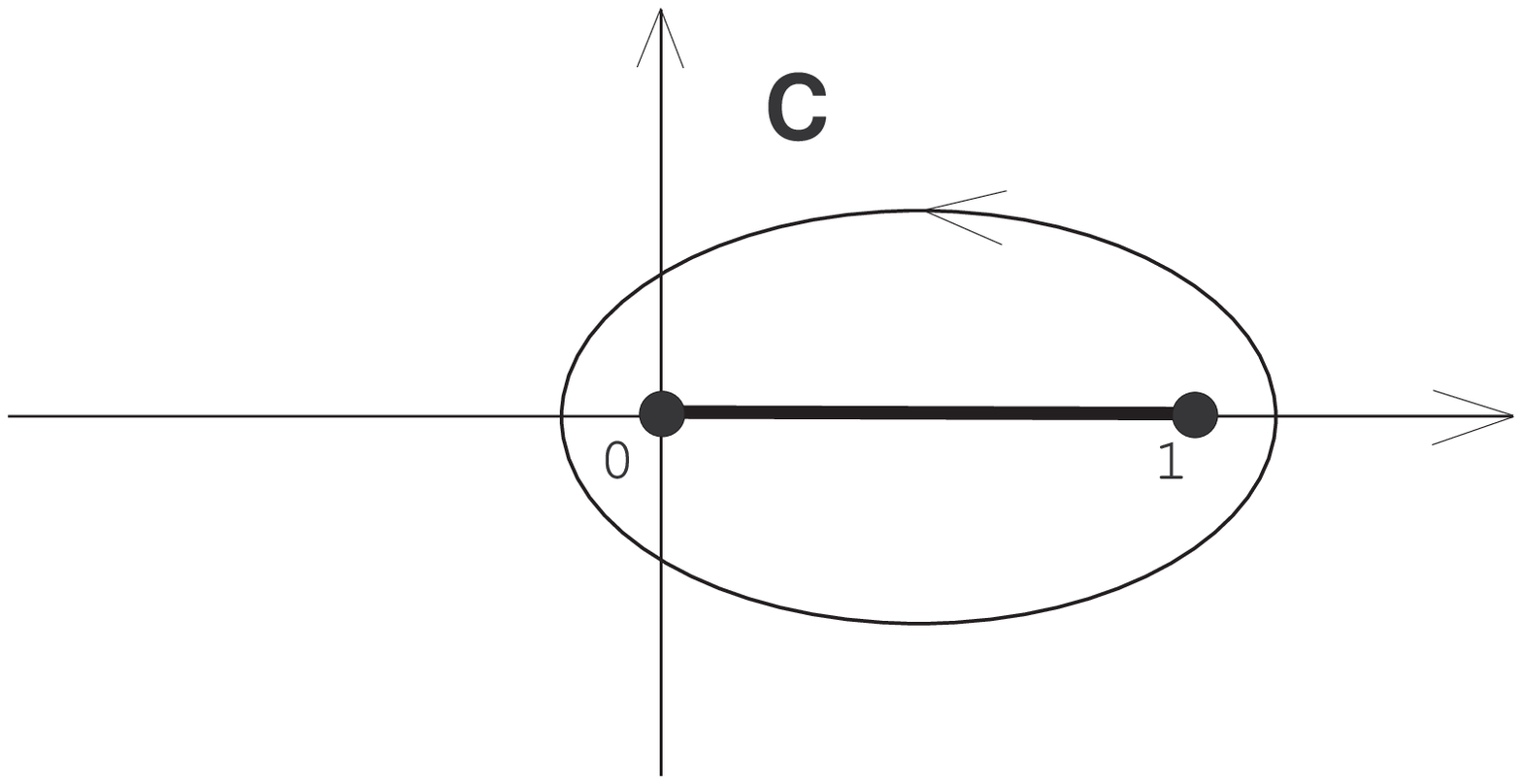}
\vskip 0.2cm {\bf Fig.~3.} Integration contour in the complex $t$-plane.
\end{center}

\subsubsection{Determination of $H[PO]$}

Consider first $H[PO]$. (28) and (30) yield
\begin{equation}\label{31}
\begin{array}{rl}
2i\lambda_0H[PO]=&\ds\frac{z}{2\pi i}\sum\limits_{n=1}^{\infty}\sum
\limits_{m=1}^{\infty}
\frac{x^ny^m}{\Gamma(n+m+1)}\,\oint\limits_{C}\,dt\,
  t^{-n}(t-1)^{n-1}(1-tz)^{m-1}
  \\
   =&\ds\frac{z}{2\pi i}\oint\limits_{C}\,\frac{dt}{(t-1)(1-zt)}
   \sum\limits_{n=1}^{\infty}\sum\limits_{m=1}^{\infty}
\frac{X^nY^m}{\Gamma(n+m+1)}\ ,
\end{array}
\end{equation}
where we have introduced the $t$-dependent variables
\begin{equation}\label{32}
X=x(t-1)/t\ ,\quad Y=y(1-zt)\ .
\end{equation}
The sum over $n$ and $m$ is easily calculated (see Appendix C) and
one obtains

\begin{equation}\label{33}
S(X,Y) = \sum\limits_{n=1}^{\infty}\sum\limits_{m=1}^{\infty}
\frac{X^nY^m}{\Gamma(n+m+1)}=1+\frac{X}{Y-X}\ e^Y-\frac{Y}{Y-X}\ e^X\ .
\end{equation}
Thus (31) splits into three integrals
\begin{equation}\label{34}
\begin{array}{rl}
2i\lambda_0H[PO]=&\ds\frac{z}{2\pi
i}\oint\limits_{C}\,\frac{dt}{(t-1)(1-zt)} \Bigl(1+ \frac{X}{Y-X}\
e^Y-\frac{Y}{Y-X}\ e^X\Bigr)\\ \equiv &I_1+I_2+I_3\ .
\end{array}
\end{equation}
These integrals are easily calculated. There is only one pole at
$t=1$ for the first integral, and no pole inside the contour for the
second one. Thus,
\begin{equation}\label{35}
I_{1}={z\over 1-z}\ ,\quad I_{2}= 0\ .
\end{equation}
The integrand in the third integral
$$I_{3}= -\frac{z}{2\pi i}\oint\limits_{C}\,\frac{dt}{(t-1)}\cdot \frac{ty
e^{x(t-1)/t}}{ty(1-tz)+x(1-t)}$$
has poles at $t=1$ and $t=t_{\pm}$, where
$$t_{\pm}=\frac{1}{2yz}(-x+y\pm\sqrt{(x-y)^2+4xyz}\,)\ . $$
As in the previous case only the pole at $t=1$ is inside the
integration contour. However, in this case the integrand has an
essential singularity at $t=0$ due to $\exp(x\frac{t-1}{t})$. One
can, however, sum the residues over all the poles outside the contour
$C$ (this can be done because the integrand vanishes at $|t|\to
\infty $ faster than $1/|t|$). Thus $I_{3}$ is obtained from the sum
of the residues at $t=t_{-}$ and $t=t_{+},$ taken with opposite signs
\begin{equation}\label{36}
I_{3} = \displaystyle
 -{ \frac{1}{(t_{+}-t_{-})}\Bigl(\frac{e^{u_+}}{1-1/t_{+}}-
   \frac{e^{u_-}}{1-1/t_{-}}\Bigr)} \ ,
\end{equation}
defining
$$ u_\pm =\ {1\over 2}\bigg( x+y\pm \sqrt{(x-y)^2+4xyz}\bigg)\ .$$
Collecting (35) and (36), we finally get
\begin{equation}\label{37}
2i\lambda_0H[PO]=\displaystyle
{z\over 1-z} +\frac{zxy}{\sqrt{(x-y)^2+4xyz}}
  \bigg( {e^{u_+}\over u_+}-{e^{u_-}\over u_-} \bigg)\ .
\end{equation}

\subsubsection{Determination of $H[PP]$}

Consider now the case of $H[PP]$.
The contribution of all diagrams with the Pomerons to the left and
right ends of each diagram has the form (27)
\begin{equation}\label{38}
\begin{array}{rll}
\ds 2i \lambda_+H[PP]=&
 \ds z\ \sum\limits_{n=2}^{\infty}\sum\limits_{m=1}^{\infty}
{x^n y^m\over(n+m)!}\ \frac{1}{m}\frac{d}{dz} \ _2F_1(-m,1-n;1;z) \
+\ e^x-x-1 \\ \equiv & \ds e^x-x-1 + I_{PP} \ .
\end{array}
\end{equation}
Repeating the arguments given for $H[PO]$, we can rewrite
$I_{PP}$ as
\begin{equation}\label{39}
\begin{array}{lll}
I_{PP} & \displaystyle{
 =\frac{z}{2\pi i}\sum\limits_{n=2}^\infty
  \sum\limits_{m=1}^{\infty}\frac{x^ny^m}{(n+m)!}\frac{1}{m}\frac{d}{dz}\
  \oint\limits_{C}\,dt\,t^{-n}(t-1)^{n-1}(1-tz)^{m}}\\
& \displaystyle{
   =-\frac{z}{2\pi i}\oint\limits_{C}\,dt\,\frac{t}{(t-1)(1-tz)}\sum\limits_
  {n=2}^{\infty} \sum\limits_{m=1}^{\infty}\frac{X^nY^m}{(n+m)!}}
\end{array}
\end{equation}
with the same $X$, $Y$ ( see (32)) and the same contour $C$ as above (see
Fig.3). The sum over $n$ and $m$ can now be immediately obtained. One finds
$$ \sum\limits_
  {n=2}^{\infty}
  \sum\limits_{m=1}^{\infty}\frac{X^nY^m}{(n+m)!}=S(X,Y)-
  \widetilde S(X,Y)\ ,
$$
where $S(X,Y)$ was given in (33) and
\eq\label{40}
\widetilde S(X,Y)=\frac{X}{Y}(e^Y-Y-1).
\en
The integral (39) can thus be rewritten in the following form
\begin{equation}\label{41}
I_{PP}=\displaystyle{
   =-\frac{z}{2\pi i}\oint\limits_{C}\,dt\,\frac{t}{(t-1)(1-tz)}[S(X,Y)-
   \tilde S(X,Y)]\equiv  {\cal I}_{PP}-\widetilde {\cal I}_{PP}}\ .
\end{equation}
Replacing $X$ and $Y$ by their expressions (32), $\widetilde{\cal
I}_{PP}$ becomes

\eq\label{42}
\widetilde{\cal I}_{PP}=\frac{z}{2\pi
i}\oint\limits_{C}\,dt\,\frac{x/y}{(1-tz)^2}\Bigl(e^{y(1-tz)}
-y(1-tz)-1\Bigr)\ ,
\en
which is zero, because there are no singularities inside the
integration contour. Now we examine ${\cal I}_{PP}$, the first
integral in (41); the only difference with the previous $[PO]$ case
(34) is the factor $t$ in the integrand and we may write
\begin{equation}\label{43}
\begin{array}{rll}
{\cal I}_{PP}=&-\frac{z}{2\pi
i}\oint\limits_{C}\,dt\,\frac{t}{(t-1)(1-tz)}\Bigl[1+\frac{X}{Y-X}e^Y-
\frac{Y}{Y-X}e^X\Bigr]  \\
\equiv &{\cal I}_{1}+{\cal I}_{2}+{\cal I}_{3}.
\end{array}
\end{equation}
We see immediately that the first and the second terms are
\begin{equation}\label{44}
{\cal I}_{1}=-\frac{z}{1-z}\ ,\quad
{\cal I}_{2}=0\ ,
\end{equation}
while the third term can be rewritten as
\begin{equation}\label{45}
 {\cal I}_{3}=
\frac{zy}{2\pi
i}\oint\limits_{C}\,dt\,\frac{t^2e^{x(t-1)/t}}{(t-1)[ty(1-tz)-x(t-1)]}\ .
\end{equation}
We have always an essential singular point at $t=0$ because of the
exponential. At the same time there are three poles outside the
contour, namely, at $t=t_{-},\ t=t_{+}$ (defined in the previous
subsection for $[PO]$) and at infinity. Thus, we can expand the
contour $C$ at very large distances from $t=0$, take the residues at
$t=t_{-}$ and at $t=t_{+}$ (with minus sign ) and write
$$
{\cal I}_{3}=-zy\,{\rm Res}_{t=t_{-}}f(t)-zy\,{\rm Res}_{t=t_{+}}f(t)
+\frac{zy}{2\pi i}\oint\limits_{R}\,dt\,f_{as}(t)\ .
$$
Here $R$ is a
circle of large radius and $f(t)$ is the integrand in (45)
$$
f(t)=\frac{t^2e^{x(t-1)/t}}{(t-1)[ty(1-tz)-x(t-1)]}\ .
$$ On the circle of large radius we can approximate $f(t)$
by its asymptotic form for $|t|\to \infty$ $\it i.e.$
$f_{as}(t)=-e^x/tyz$. Thus
$$
\frac{zy}{2\pi
i}\oint\limits_{R}\,dt\,f_{as}(t)=-\frac{e^x}{2\pi
i}\oint\limits_{R}\,\frac{dt}{t}=-e^x.
$$
Collecting this integral
and the residues at $t=t_{-},\ t=t_{+}$ we obtain for ${\cal I}_{3}$ the
following expression~\footnote{
It is possible, of course, to obtain this result more exactly replacing
the integration variable in (45) by $1/t$. After this
transformation the singularities under interest are inside a new
integration contour.}
\begin{equation}\label{46}
{\cal I}_{3}=-e^x+\frac{t_{-}^2e^{x-x/t_{-}}}{(t_{-}-1)(t_{-}-t_{+})}+
\frac{t_{+}^2e^{x-x/t_{+}}}{(t_{+}-1)(t_{+}-t_{-})}\ .
\end{equation}
Taking into account (38) and the expressions for $t_{\pm}$ we
obtain the final expression for $H[PP]$
\begin{equation}\label{47}
  \begin{array}{rll}
    2i\lambda_+H[PP]& =-x-1-\ds\frac{z}{1-z}-
    \frac{\ x}{2\sqrt{(x-y)^2+4xyz}} \\
    &\times   \ds\Bigg\{
     \bigg({-x+y-\sqrt{(x-y)^2+4xyz}}\bigg)
 \   {e^{u_+}\over u_+}  \\
                    & -\ds\bigg({-x+y+\sqrt{(x-y)^2+4xyz}}\bigg)
 \    { e^{u_-}\over u_-}\Bigg\}  \ .
  \end{array}
\end{equation}

\subsubsection{Full eikonalized amplitude}

The full eikonalized amplitude (12), is now obtained by adding the
two rescattering contributions found above to the Born amplitude and
using the substitution $x \longleftrightarrow y$ to get
$\lambda_-H[OO]$ from $\lambda_+H[PP]$. Putting everything together,
we finally have the three-parameters eikonalized amplitudes for the
$\bar pp$ and $pp$ elastic scattering process
\begin{equation}\label{48}
\begin{array}{ll}
H^{\bar pp}_{pp,GE}& (s,b) =
            \ds{  {i\over 2 (\lambda_0^2-\lambda_-\lambda_+)}  } \\
&\times\bigg\{    a + e^{ \ds i(\lambda_+ h_+ \pm \lambda_- h_-)} \
\displaystyle \big[-a \cos{\phi_\pm} +
i {c_+h_+\pm c_-h_-\over\phi_\pm} \sin{\phi_\pm}\big]  \bigg\}  \  ,
\end{array}
\end{equation}
where we have introduced three constants $a$ and $c_{\pm}$ defined as
\begin{equation}\label{49}
a=2\lambda_0-\lambda_+-\lambda_-\ ,
\end{equation}
\begin{equation}\label{50}
c_\pm=
\lambda_+\lambda_--2\lambda_0^2-\lambda_\pm^2+2\lambda_0\lambda_\pm\ ,
\end{equation}
in terms of the parameters of the model and two functions (of $s$ and $b$)

\begin{equation}\label{51}
\phi_\pm=\sqrt{(\lambda_+h_+ \mp \lambda_-h_-)^2\ \pm 4\lambda_0^2h_+
h_-}\ .
\end{equation}

\subsection{Particular cases}

After finding general expressions for the
amplitudes, valid at any $z$, we investigate some particular cases

\medskip
\hskip 1.cm {\bf (i)} $z=1$ or $\lambda_0^2=\lambda_+\lambda_-$.
\medskip

\noindent We may use the results of Sect.~3.1 to check the results of
Sect.~2, relative to the two parameters parameterization. In fact
(9)-(13) directly follow from (27),(28) if $_2F_1$ is calculated at
$z=1$ using (30).

\medskip
\hskip 1.cm {\bf (ii)} $z=1-\varepsilon$, $|\varepsilon|\ll 1 $.
\medskip

It is instructive to derive an expansion of the general expressions
(37) and (47) in terms of the small quantity $\varepsilon = 1-z$.
Consider for example (37) for $H[PO]$. One can see that the first
term ${z \over 1-z}$ and the third one with $\exp(u_-)/u_{-}$ have
singularities at $ \varepsilon = 0$ which cancel each other. Keeping
the zeroth and first powers of $\varepsilon$ we obtain the following
result for $H[PO]$ at $z\approx 1$
\begin{equation}\label{52}
\begin{array}{ll}
  2i&\lambda_0 H[PO]\simeq\displaystyle{\frac{xy}{(x+y)^2}\Bigg\{
  e^{x+y}-x-y-1-\varepsilon\Bigl[-x-y-1}\\&+
  \displaystyle{\frac{xy}{2(x+y)^2}(
  (x+y)^2+4(x+y)+6)+ e^{x+y} [ 1+{xy\over x+y}-{3xy\over (x+y)^2} ]
  \Bigr]\Bigg\}  }\ .
\end{array}
\end{equation}
Similarly the singularities at $\varepsilon=0$ cancel in $H[PP]$ and
we obtain
\begin{equation}\label{53}
\begin{array}{ll}
  2i&\lambda_+H[PP]\simeq\displaystyle{\frac{x^2}{(x+y)^2}\Bigg\{
  e^{x+y}-x-y-1+\varepsilon\frac{y}{(x+y)^2}}\\&\times
  \displaystyle{\Bigl[\frac{1}{2}(x+y)
(y^2+xy+2y-2x-4)+3y+e^{x+y}(2x-y-x^2-xy)\Bigr] \Bigg\}}\ .
\end{array}
\end{equation}
Using (12) we obtain the full eikonalized amplitude in the
$b$-representation. Of course at $\varepsilon=0$ (or $z=1$) these
results coincide with the similar expressions of the Sect.2.

\medskip
\hskip 1.cm {\bf (iii)} $z\ll 1$ or $\lambda_0^2\ll
\lambda_+\lambda_-$ (weak coupling limit).
\medskip

\noindent This case is of conceptual interest because it corresponds
to a clear separation between the Pomeron contribution and the
Odderon one. As a result of this approximation, we obtain eikonalized
Pomeron, eikonalized Odderon and small interference terms
proportional respectively to $z/\lambda_0 $, $z/\lambda_+$,
$z/\lambda_-$. This is illustrated in Fig.~4. \vskip 0.4cm
\begin{center}
\includegraphics*[scale=0.6]{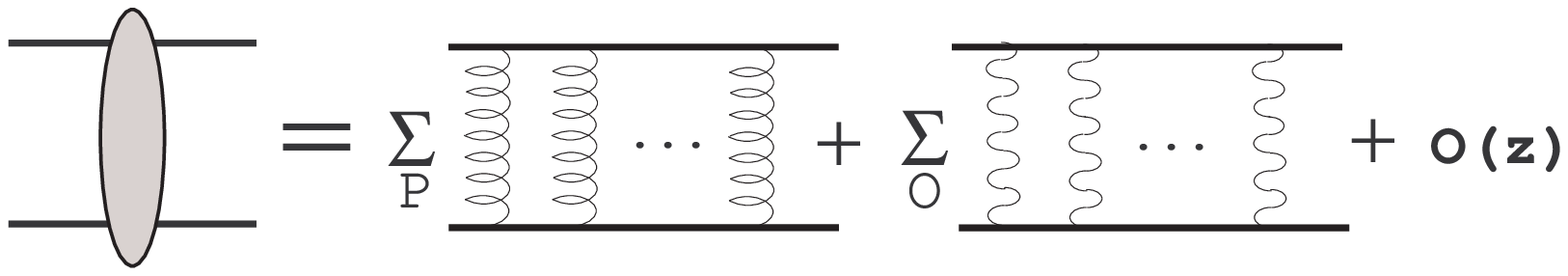}
\vskip 0.3cm
{\bf Fig.~4.} Diagram representation of $H(s,b)$ at $z\ll 1$.
\end{center}
 The expressions for $H^{\bar p p}_{pp}$ can be easily obtained.

\medskip
\hskip 1.cm {\bf (iv)} $z\gg 1$ ( strong coupling limit).
This case is the more complicate one to obtain approximate expressions for
$H^{\bar p p}_{pp}$ in closed analytical form and we will not give here the
formulae that follow. Preliminary numerical investigations, however, seem to
indicate that this is the choice that gives the best account of the data.

\bigskip
\subsection{Unitarity constraints}

We restrict our considerations to the case
\begin{equation}\label{54}
|h_{-}(s,b)| \ll |h_{+}(s,b)|\ .
\end{equation}
which is the correct assumption in the high energy region (at least
for $0\leq b^2 < R^2)$) when the Pomeron dominates over the Odderon
and other Reggeons (see Appendix B). If we expand the full
eikonalized amplitude (48) in terms of the small complex quantity
$\epsilon={h_{-}(s,b) \over h_{+}(s,b)}$ we have
\begin{equation}\label{55}
\phi_\pm \approx \lambda_{+}h_{+}(1\mp
\epsilon({\lambda_{-} \over
\lambda_{+}}-2{\lambda_{0}^2 \over\lambda_{+}^2})) \ ,
\end{equation}
\begin{equation}\label{56}
{c_+h_+ \pm c_-h_-\over \phi_\pm}\approx {c_+\over\lambda_+}
\left[1\pm \epsilon ({c_-\over c_+}+{\lambda_-\over\lambda_+}-
2{\lambda_0^2\over\lambda_+^2}
)\ \right]\ ,
\end{equation}
\begin{equation}\label{57}
H^{\bar pp}_{pp}
         \approx \displaystyle {i\over 2 (\lambda_0^2-\lambda_+\lambda_-)}
\bigg\{a +  \displaystyle {1\over 2} \big(-a+{c_+\over \lambda_+}\big)
e^{2i(\lambda_+h_+\pm h_{-}{\lambda_0^2\over \lambda_+})}-
{1\over 2}\big(a+{c_+\over \lambda_+}\big)e^{\pm 2i\lambda_-h_-
\big(1-{\lambda_0^2\over \lambda_+\lambda_-}\big)}\bigg\} \ .
\end{equation}
The second term in the last expression can be neglected because $h_+(s,b)$
is mainly imaginary and $|h_+|\to \infty$ when $s\to \infty$
if $0\leq b^2 < R^2$ (see Appendix B).
We obtain
\begin{equation}\label{58}
H^{\bar pp}_{pp}
    \approx\displaystyle {i\over 2 (\lambda_0^2-\lambda_+\lambda_-)}
\bigg\{ a-{1\over 2}(a+{c_+\over\lambda_+}) e^{\pm
2i\lambda_-h_-(1-z)} \bigg\} \ .
\end{equation}
The following comments apply.

{\bf 1.} If $\delta_{-}>0$, the unitarity inequality
\begin{equation}\label{59}
  |H_{pp}^{\bar pp}(s,b)|\leq 1
\end{equation}
cannot be satisfied unless the factor in front of
the exponential in (58) is zero since
$|\exp(\pm 2ih_-\lambda_-(1-z))|\to \infty$ at $s\to \infty$ either
for $pp$ or for $\bar pp$. Thus, in general, the only possibility to
preserve
the unitarity restriction is $\delta_{-}\leq 0.$ In this case
$|h_-|\to 0$ when $s\to \infty$ and $0\leq b^2 <R^2$ (see Appendix B) and
consequently
\begin{equation}\label{60}
 H_{pp}^{\bar pp}(s,b)\approx \displaystyle
 {\frac{i}{2}\Bigg\{\frac{2\lambda_0-\lambda_+-\lambda_-}
 {\lambda_+\lambda_-(z-1)}+\frac{(\lambda_+-\lambda_0)^2}
 {\lambda_+^2\lambda_-(z-1)}\Bigg\} =\frac{i}{2\lambda_+}}.
\end{equation}
As a consequence, if $\delta_{-}\leq 0$ the restriction
\begin{equation}\label{61}
\lambda_+\geq 1/2
\end{equation}
\noindent follows from the unitarity constraint. Let us emphasize that
there are no restrictions on $\lambda_-$ and $\lambda_0$ so long as
$z=\lambda_0^2/\lambda_-\lambda_+\neq 1.$

{\bf 2}. If
\begin{equation}\label{62}
  \lambda_0=\lambda_+.
\end{equation}
however, the factor in front of the exponential in (58) vanishes
 and the amplitude can be approximated as
\begin{equation}\label{63}
 H_{pp}^{\bar pp}(s,b)\approx \displaystyle
 {\frac{i}{2}\Bigg\{\frac{2\lambda_0-\lambda_+-\lambda_-}
 {\lambda_+\lambda_-(z-1)}\Bigg\}=\frac{i}{2\lambda_+}}
\end{equation}
and we find again that (61) is valid irrespective of the sign of
$\delta_-$. Finally, if $\alpha_{+}(0)> \alpha_{-}(0)$ (as
always assumed to be the case), the condition (61) can be satisfied
if
\begin{equation}\label{64}
\lambda_0=\lambda_+\neq \lambda_- \ .
\end{equation}

Table 2 collects the results found here
and in subsection  2.2 concerning the unitarization constraints
on the parameters assuming Pomeron dominance,
\ie when
 $\sigma^{pp,p\bar p}_{tot}\propto\alpha'_+\delta_+\ell n^2 s $.

\def\init{\tabskip 0pt\offinterlineskip}
\def\crr{\cr\noalign{\hrule}}

$$\vbox{\init\halign to 14.cm{
\strut#&\vrule#\tabskip=1em plus 2em&

\hfil$#$\hfil&
\vrule$\,$\vrule#&
\hfil$#$\hfil&
\vrule#&
\hfil$#$\hfil&
\vrule#&
\hfil$#$\hfil&
\vrule#&
\hfil$#$\hfil&
\vrule#\tabskip 0pt\crr
& & &&  z=1 \, {\rm (QE)} && z=1  \, {\rm (GE)} &&
z\neq 1  \, {\rm (GE)} &&  z\ne 1 \, {\rm (GE)} &\crr
&&\lambda_+&& \geq 1/2 &&  \geq1/2  &&  \geq 1/2   && \geq 1/2     &\cr
&&\lambda_-&&=\lambda_+&&{\rm  any } &&{\rm  any}&&{\rm any}&\cr
&&\lambda_0&&=\lambda_+&&=\sqrt{\lambda_+\lambda_-}&&=\lambda_
+&&{\rm any}&\crr
&&\delta_+ &&\geq\delta_-&& >0   && \geq \delta_-&& >0  &\cr
&&\delta_- && >0 &&\leq 0&&   >0\quad {\rm  or}\quad < 0  && \leq 0  &\crr
}}$$

\medskip
\noindent \centerline {\bf Table 2 } Summary of the results of
subsections 2.2, 3.4 for 4 classes of eikonalized models dominated by
the Pomeron, respecting unitarity constraints.


\bigskip
\section{\large{\bf Conclusion}}
First of all, let us remind that we have considered only an
eikonalization procedure rather than a complete unitarization and
that we have analyzed only the case of elastic scattering. As
mentioned at the end of Sect.~2.1, the case of diffraction
dissociation would imply the insertion of new couplings. With these
new effective couplings, the main difference of the diffraction
dissociation amplitudes derived in analogy with the elastic ones
would be in the energy independent parts of their slopes (which, in
fact, is experimentally quite different). Asymptotically, these more
general amplitudes have similar behaviors and the results issued from
unitarity would remain the same.

A recent controversy concerns the sign of $\delta_{-}=\delta_O$ ({\it
i.e.} of the difference with unity of the Odderon intercept at
$t=0$). This quantity, (for which indications have been found long
ago [9] that it should be negative) was initially \cite{gln} believed
to be positive from QCD calculations of the Odderon trajectory, but
counterarguments where then given [17] that $\delta_{-}$ should
actually be negative. More recently [18], this parameters has been
calculated and found to be negative~\footnote{Latest QCD
calculation~[19] gives $\delta_O=0$.}. On purely phenomenological
grounds, but with absolute rigor, we can state, in the more general
case of Sect.3, that $\delta_{-}$ {\it must} be negative or null,
unless the specific equality $\lambda_+ = \lambda_0$ holds in which
case this sign can also be positive. Of course, nothing prevents {\it
a priori} such an equality between quantities related to the coupling
of particles to the Pomeron and the Odderon to hold but it certainly
looks like a rather peculiar relation and should it turn out to be
indeed valid, it certainly would deserve further investigation to
understand its implications.

The main merit of our paper, however, lies in the great generality of
the formalism we have developed leading to a complete 3-parameters
eikonalization. In spite of the apparent complications introduced, it
is quite likely that these are not of a purely abstract interest and,
indeed, we are already exploring its phenomenological implications in
describing all elastic $pp$ and $\bar pp$ data simultaneously [8]. The
standard eikonal, in fact, is not suitable for a realistic physical
description of the elastic amplitude where two classes of Reggeons
contributions have to be kept: {\it e.g.} the Pomeron (essential to
describe the small $|t|$ domain) and the Odderon (essential to
account for the large $|t|$ domain). This is especially visible when
discussing the $\bar pp$ and the
$pp$ scattering at high energy. In fact, in this case, as we have
seen, it may be appropriate to group together the crossing even and
the crossing odd combinations $P+f$ and $O+\omega$ but
one could also argue that it is not necessary to eikonalize the secondary
Reggeons (because they do not imply any violation of unitarity)
in which case one would interpret the crossing
even and the crossing odd parts as due simply to the Pomeron and the
Odderon.

We believe that the present generalization can be successfully applied
to a phenomenological description of all high energy $pp$ and $\bar pp$
elastic scattering data where both the Pomeron and the Odderon contribute.
This is under investigation presently [8].

\bigskip

\noindent
{\bf Acknowledgements}

\noindent We thank Dr. A. Bugrij for very fruitful discussions. One
of us, E. M. would like to thank the Theory Groups of the
Universities of Lyon and Torino for their kind invitations. Financial
support from the INFN and the MURST of Italy is gratefully
acknowledged. The authors would also like to thank the Referee for a
very useful remark which has led to adding the last sentence of
Section 2.1 and the first of the Conclusion.

\bigskip

\bigskip
\bigskip

\centerline {\bf APPENDIX A}
\medskip

\centerline{\bf Two-parameters amplitude}
\centerline {\bf Proof of (13)}

\bigskip

We consider for example (9) and rewrite
$$
H[PP]={1\over
2i\lambda_+}\ \phi_{PP}(x,y)\ ,
$$
defining $x=2i \lambda_+h_+$,
$y=2i \lambda_-h_-$ and
$$
\phi_{PP}(x,y)\ =\
\sum\limits_{k=0}^{\infty}
\sum\limits_{m=0}^{\infty}\frac{x^{k+2}y^m}{k!m!(k+m+2)(k+m+1)}\ .
$$

After the substitution $y=xu$ this equation becomes
$$
\phi_{PP}(x,y)\
=\ \widetilde\phi_{PP}(x,u)\ =\ \sum\limits_{k=0}^{\infty}
\sum\limits_{m=0}^{\infty}\frac{x^{k+m+2}u^m}{k!m!(k+m+2)(k+m+1)}\,
$$
which satisfies
$$
 \widetilde\phi_{PP}(x,0)\ =\
{\partial\widetilde\phi_{PP}\over\partial x}(x,0)\ =\ 0\quad {\rm
and}\quad  {\partial^2\widetilde\phi_{PP}\over\partial x^2}(x,u)\
=\ e^{x+xu}\ .
$$
Its integration leads to
$$
\phi_{PP}(x,y)\ =\
x^2\varphi (x,y)\quad {\rm with }\quad \varphi(x,y)=\frac{
(e^{x+y}-1)}{(x+y)^2}- \frac{1}{x+y}       \ .
$$
Finally we obtain
$$
H[PP]={x^2\over 2i\lambda_+}\ \varphi(x,y)\ .
$$
Similarly
$$
\begin{array}{cl}
  H[OO]= & \ds{y^2\over 2i\lambda_-}\ \varphi(x,y)\ , \\
  H[PO]= & \ds\frac{xy}{2i\sqrt{\lambda_+\lambda_-}}\ \varphi(x,y)\ .
\end{array}
$$
Collecting these results, we derive the final expression (13) for the
impact parameter eikonalized amplitude in the two-parameters GE procedure.

\bigskip

\bigskip
\centerline {\bf APPENDIX B}

\medskip

\centerline {\bf Born Pomeron and Odderon amplitudes}
\bigskip

In this Appendix we give the expressions of the amplitudes which we
use to discuss the unitarity constraints. For simplicity, we consider
the specific model known as Regge monopole (see below) but, as we
will argue, the conclusions are quite model independent. For the same
reason (simplicity), the
 input crossing even and odd {\it (Born)} amplitudes entering in (1) are
 approximated, at high energy, by the Pomeron and Odderon contributions,
 respectively
$$
a_+(s,t)=\ a_P(s,t)\ ,\quad
a_-(s,t)=\ a_O(s,t)\ .
$$

These amplitudes (in the ($s,t$)-representation) are
$$
a_{\pm}(s,t)= m_{\pm} \ \tilde s^{\alpha_{\pm}(t)}\ e^{b_\pm t} \ ,
$$
for the monopole. In this equation,
$$
\tilde s\ =\ {s\over s_0} \ e^{-i{\pi\over 2}}\ ,\quad
(s_0=1\ {\rm GeV}^2)
$$
to respect $s-u$ crossing and linear trajectories $\alpha_{\pm}(t)$ are
considered
$$\alpha_{\pm}(t)= 1+\delta_{\pm}+\alpha_{\pm}'t\ .
$$
The well known unitarity constraints set the Pomeron-Odderon hierarchy
$$ \alpha'_{+} > \alpha'_{-}\ ,\quad
\delta_{+} > \delta_{-}\ ,\quad {\rm with}\ \delta_{+} >0 . $$ With
the above choices, the "coupling" $m_{+}$ is real and negative while
$m_{-}$ is purely imaginary. The optical theorem sets the
normalization $\sigma_{tot}= {4\pi\over s} \Imag {A(s,t=0)}$.

The corresponding amplitudes $h_{\pm}$ (in the ($s,b$)-representation)
are readily obtained by the Fourier-Bessel's transform (3) and they are
$$ h_{\pm}(s,b)= {m_{\pm} \widetilde {s}^{\alpha_{\pm}(0)}\over
{4sM_{\pm}}} \ {e^{-b^2/4M_{\pm}}}\ , \quad {\rm with} \
M_{\pm}= \alpha'_{\pm} \ln  \tilde s+ b_{\pm}\ .       $$

It is easy to show that, for large s values and $0\leq b^2 < R^2$,
$$|e^{2i\lambda_{+}h_{+}}| \approx
\mbox{exp}\left({m_{+}\lambda_{+}\cos({\pi \over 2}\delta_{+})
 \over 2s_0\alpha'_{+}\ln s}(\frac{s}{s_0})^{\delta_{+}}\right) \ .$$
Due to the negative sign of $m_{+}$ and to the positive sign of
$\delta_{+}$, this quantity goes to zero when $ s \to  \infty$ (for both
processes -$\bar pp$ and $pp$-).

Similarly, for large s values and $0\leq b^2 < R^2$,
$$|e^{{\pm}2i\lambda_{-}h_{-}(1-z)}| \approx
\mbox{exp}\left({\pm}{\widetilde{m}_{-}(\lambda_{+}\lambda_{-}-
\lambda^{2}_0)
\over 2s_0\alpha'_{-}\lambda_{+}\ln s} \sin({\pi \over
2}\delta_{-})(\frac{s}{s_0})^{\delta_{-}}\right)\ ,
$$
(recall that the sign $\pm$ distinguishes the $\bar pp$ and $pp$
scatterings) where we have defined $m_{-}=i\widetilde{m}_{-}$. One of
these quantities diverges if $\delta_{-}>0$, irrespective the sign of
$\widetilde{m}_{-} (\lambda_{+}\lambda_{-}-\lambda_0^2)$. Thus,
$\delta_-$ must be negative or null (see subsect 3.4).

Had we chosen a dipole \footnote {The difference between a monopole
and a dipole is essentially that the amplitude of the second grows
with an additional power of $\ln s$ .} instead of a monopole
$$
a_{\pm}(s,t)= d_{\pm} \ \tilde s^{\alpha_{\pm}(t)}\
e^{b_{\pm}(\alpha_{\pm}(t)-1)}
(b_{\pm}+ \ln \tilde s)\ ,
$$
(or a combination of a monopole with a dipole), similar conclusions would
follow.

\bigskip

\bigskip
\centerline {\bf APPENDIX C}

\medskip

\centerline{\bf Proof of equation (33)}

\bigskip

We start from the definition
$$
S(x,y) = \sum\limits_{n=1}^{\infty}\sum\limits_{m=1}^{\infty} \frac{x^n
y^m}{(n+m)!}\ =\ xy \ \sum\limits_{q=0}^{\infty}
{y^q\over\Gamma(q+3)}
\sum\limits_{p=0}^{\infty}{x^p\over(q+3)_ p}\ ,
$$
or~\cite{pru,ha}
$$
S(x,y) = {xy\over 2} \sum\limits_{q=0}^{\infty}
{y^q\over (3)_q}\ _1F_1(1;q+3;x)\ $$
$$ ={y\over x}\int_0^x dt\, e^{ty/x} (x-t) _1F_1(1;2;x-t)=
{y\over x}\int_0^x dt\, \bigg(e^x\cdot e^{t(y/x-1)}-e^{ty/x}\bigg) $$
$$=\ \left[{x\over
y-x}e^y+{y\over x-y}e^x+1\right]\ = 1+ {{x e^y - y e^x} \over {y-x}}.
$$

\end{document}